\documentclass[twoside,twocolumn,9pt]{article}
\usepackage{extsizes}
\usepackage[super,sort&compress,comma]{natbib} 
\usepackage[version=3]{mhchem}
\usepackage[left=1.5cm, right=1.5cm, top=1.785cm, bottom=2.0cm]{geometry}
\usepackage{balance}
\usepackage{mathptmx}
\usepackage{multirow}
\usepackage{makecell}
\usepackage{sectsty}
\usepackage{graphicx} 
\usepackage{lastpage}
\usepackage[format=plain,justification=justified,singlelinecheck=false,font={stretch=1.125,small,sf},labelfont=bf,labelsep=space]{caption}
\usepackage{float}
\usepackage{fancyhdr}
\usepackage{fnpos}
\usepackage[english]{babel}
\addto{\captionsenglish}{%
  
}
\usepackage{array}
\usepackage{droidsans}
\usepackage{charter}
\usepackage[T1]{fontenc}
\usepackage[usenames,dvipsnames]{xcolor}
\usepackage{setspace}
\usepackage[compact]{titlesec}
\usepackage{hyperref}

\definecolor{cream}{RGB}{222,217,201}

\begin{document}

\pagestyle{fancy}
\thispagestyle{plain}
\fancypagestyle{plain}{
\renewcommand{\headrulewidth}{0pt}
}

\makeFNbottom
\makeatletter
\renewcommand\LARGE{\@setfontsize\LARGE{15pt}{17}}
\renewcommand\Large{\@setfontsize\Large{12pt}{14}}
\renewcommand\large{\@setfontsize\large{10pt}{12}}
\renewcommand\footnotesize{\@setfontsize\footnotesize{7pt}{10}}
\makeatother

\renewcommand{\thefootnote}{\fnsymbol{footnote}}
\renewcommand\footnoterule{\vspace*{1pt}%
\color{cream}\hrule width 3.5in height 0.4pt \color{black}\vspace*{5pt}} 
\setcounter{secnumdepth}{5}

\makeatletter 
\renewcommand\@biblabel[1]{#1}            
\renewcommand\@makefntext[1]%
{\noindent\makebox[0pt][r]{\@thefnmark\,}#1}
\makeatother 
\renewcommand{\figurename}{\small{Fig.}~}
\sectionfont{\sffamily\Large}
\subsectionfont{\normalsize}
\subsubsectionfont{\bf}
\setstretch{1.125} 
\setlength{\skip\footins}{0.8cm}
\setlength{\footnotesep}{0.25cm}
\setlength{\jot}{10pt}
\titlespacing*{\section}{0pt}{4pt}{4pt}
\titlespacing*{\subsection}{0pt}{15pt}{1pt}

\fancyfoot{}
\fancyfoot[LO,RE]{\vspace{-7.1pt}\includegraphics[height=9pt]{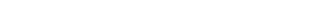}}
\fancyfoot[CO]{\vspace{-7.1pt}\hspace{13.2cm}\includegraphics{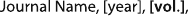}}
\fancyfoot[CE]{\vspace{-7.2pt}\hspace{-14.2cm}\includegraphics{head_foot/RF}}
\fancyfoot[RO]{\footnotesize{\sffamily{1--\pageref{LastPage} ~\textbar  \hspace{2pt}\thepage}}}
\fancyfoot[LE]{\footnotesize{\sffamily{\thepage~\textbar\hspace{3.45cm} 1--\pageref{LastPage}}}}
\fancyhead{}
\renewcommand{\headrulewidth}{0pt} 
\renewcommand{\footrulewidth}{0pt}
\setlength{\arrayrulewidth}{1pt}
\setlength{\columnsep}{6.5mm}
\setlength\bibsep{1pt}

\makeatletter 
\newlength{\figrulesep} 
\setlength{\figrulesep}{0.5\textfloatsep} 

\newcommand{\topfigrule}{\vspace*{-1pt}%
\noindent{\color{cream}\rule[-\figrulesep]{\columnwidth}{1.5pt}} }

\newcommand{\botfigrule}{\vspace*{-2pt}%
\noindent{\color{cream}\rule[\figrulesep]{\columnwidth}{1.5pt}} }

\newcommand{\dblfigrule}{\vspace*{-1pt}%
\noindent{\color{cream}\rule[-\figrulesep]{\textwidth}{1.5pt}} }

\makeatother

\twocolumn[
  \begin{@twocolumnfalse}
{\includegraphics[height=30pt]{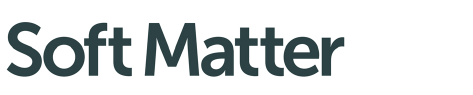}\hfill\raisebox{0pt}[0pt][0pt]{\includegraphics[height=55pt]{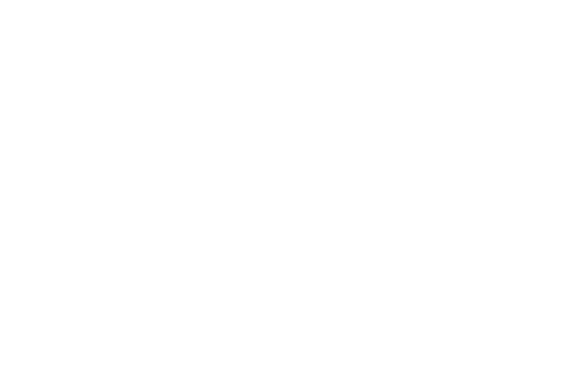}}\\[1ex]
\includegraphics[width=18.5cm]{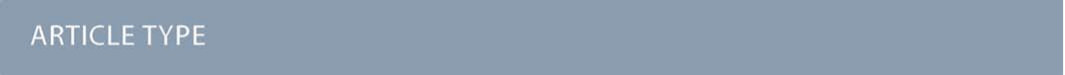}}\par
\vspace{1em}
\sffamily
\begin{tabular}{m{4.5cm} p{13.5cm} }

\includegraphics{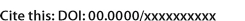} & \noindent\LARGE{\textbf{Twist and Measure: Characterizing the Effective Radius of Strings and Bundles under Twisting Contraction}} \\ 
\vspace{0.3cm} & \vspace{0.3cm} \\

 & \noindent\large{Jesse M. Hanlan,\textit{$^{a}$} Gabrielle E. Davis,\textit{$^{a,b}$} and Douglas J. Durian\textit{$^{a}$}} \\

\includegraphics{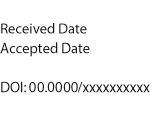} & \noindent\normalsize{We test the standard model for the length contraction of a bundle of strings under twist, and find deviation that is significantly greater than typically appreciated and that has a different nature at medium and large twist angles.  By including volume conservation, we achieve better fits to data for single-, double-, and triple-stranded bundles of Nylon monofilament as an ideal test case. This gives a well-defined procedure for extracting an effective twist radius that characterizes contraction behavior. While our approach accounts for the observed faster-than-expected contraction up to medium twist angles, we also find that the contraction is nevertheless slower than expected at large twist angles for both Nylon monofilament bundles and several other string types. The size of this effect varies with the individual-string braid structure and with the number of strings in the bundle.  We speculate that it may be related to elastic deformation within the material. However, our first modeling attempt does not fully capture the observed behavior.}\\
\end{tabular}

\end{@twocolumnfalse} \vspace{0.6cm}

]


\renewcommand*\rmdefault{bch}\normalfont\upshape
\rmfamily
\section*{}
\vspace{-1cm}


\footnotetext{\textit{$^{a}$~Department of Physics and Astronomy, University of Pennsylvania, Philadelphia, PA 19104, USA, Email: jhanlan@sas.upenn.edu, djdurian@physics.upenn.edu}}
\footnotetext{\textit{$^{b}$~Department of Physics, University of Maryland Baltimore County, Baltimore, MD 21250, USA, Email: sx46252@umbc.edu}}





\section{Introduction}

\begin{figure}[h]
    \centering
    \includegraphics[width=8cm]{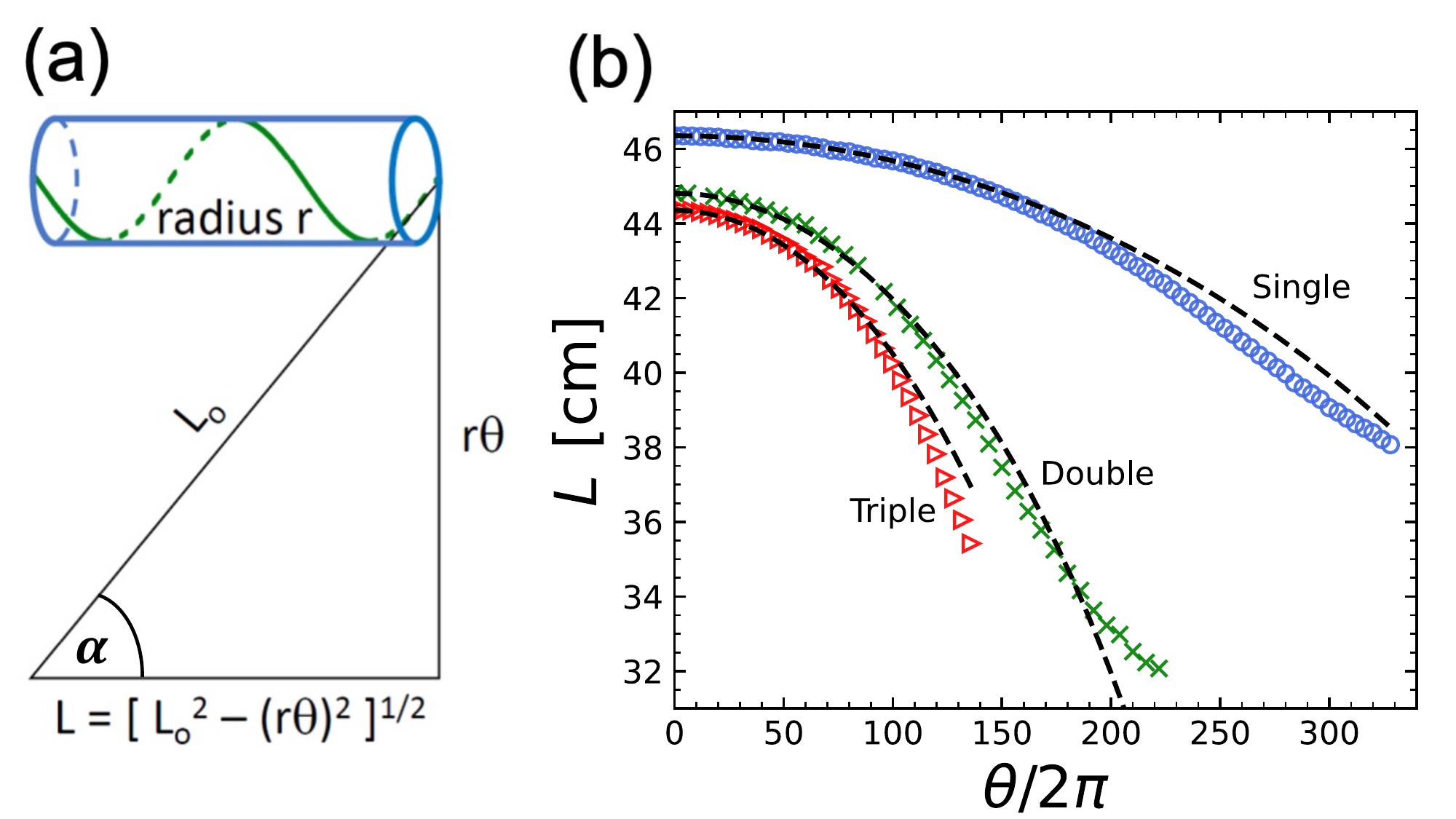}
    \caption{(a) Schematic depiction of the standard model for the length contraction of a bundle of strings, Eq.~(\ref{eq:theory}). Under twist, a straight line along an imagined cylinder of radius $r$ transforms into a helix, shown here in green. Mentally unrolling the helix through the twist angle $\theta$ forms a right triangle with the hypotenuse equal to the uncontracted length $L_0$, base equal to the contracted length $L$, and height $r\theta$. The helix angle is given by $\cos\alpha=L/L_0$. For a twisted pair, the helix is usually assumed to represent the contact line between the two strings\cite{Shoham2005}. (b) Data for contracted length $L$ versus number of full twists $n=\theta/2\pi$ for single-, double-, triple-strands of 0.41~mm diameter Nylon monofilament string. Measurement uncertainties are smaller than the markers, and thus not shown. The dotted curves represent fits to Eq.~(\ref{eq:theory}) for small angles. Note that the leading deviation, first apparent at medium angles, is faster-than-expected contraction. }
    \label{fig:theorydiagram}
\end{figure}

Taut strings are useful mechanical devices for conveying and redirecting forces in a wide range of settings, including familiar machines such as pulleys or capstans. Behavior under twist is also important in regards to the mechanics of elastic rods and bundles\cite{Mahadevan2005, Reis2013, Reis2014, Kudrolli2017, Kudrolli2018}, the friction forces that hold yarn together\cite{Goldstein2018, Crassous2022, Kudrolli2022}, and also the replication, repair, and recombination of DNA\cite{Kamien1997,Gore2006}. More specific to this work, pairs and bundles of strings are also useful based on their contraction and the related ability to transfer torque under twist. For twisted-string actuators\cite{Shoham2005, Guzek2012, Palli2013, Gaponov2013, Gaponov2014, Tavakoli2016}, as well as for ``button-on-string" or ``buzzer" toys\cite{Schlichting2010, Bhamla2017, Tang2018}, it is usually accepted that a straight length contracts with twist according to
\begin{eqnarray}
    L &=& \sqrt{L_{0}^{2} - (r\theta)^{2}}, \label{eq:theory}
\end{eqnarray}
where $L_0$ is the initial length, $r$ is an effective twist radius of the string or bundle, and $\theta$ is the axial twist angle of one end with respect to the other. This may be rationalized via Fig.~\ref{fig:theorydiagram}a, where a helix representing the twisted structure of a string or bundle is mentally unrolled, forming a right triangle with base $L$ equal to the contracted length, height $r\theta$ equal to the rolling distance, and hypotenuse $L_0$ equal to the initial uncontracted length given by the arclength of the helix -- assumed to be inextensible.  However, fits of Eq.~(\ref{eq:theory}) to data, where $r$ and/or $L_0$ are adjusted, are felt to be satisfactory\cite{Shoham2005, Wurtz2010, Guzek2012, Palli2013, Gaponov2014} even though systematic deviations are apparent and unaccounted for. Furthermore, the fitted value of $r$ in terms of geometrical parameters of the system remains unclear. In part this is because fitted values depend on the range of the data, owing to the systematic deviation. Compounding difficulties are that real strings are somewhat extensible, have a braided or core-shell structure, or are even floofy like yarn\cite{Markande2020}, so that the geometrical radius is ill-defined and exhibits complex compaction and deformation during twist.

Specific values of reported, or assumed, effective twist radii relating to Eq.~(\ref{eq:theory}) are as follows. For the case of two strands twisted together, many authors agree with the proposal by Shoham\cite{Shoham2005} that the effective twist radius equals the geometrical radius of a single strand\cite{Wurtz2010, Guzek2012, Palli2013, Gaponov2014, Tavakoli2016, Schlichting2010, Bhamla2017}. For multi-strand bundles, Guzek \textit{et al.}\cite{Guzek2012} suggest that the effective radius should be half the distance between the outermost strands, but note that many potential packings are possible for a given number of strands. Palli \textit{et al.}\cite{Palli2013} propose instead that only the outermost strands contribute to twisting, and the effective radius should be the distance from the center of the packing to the center of outermost strand. Tavakoli \textit{et al.}\cite{Tavakoli2016} suggest that as new strands are added to a bundle, they attempt to symmetrically pair on opposite sides of the packing. This gives a unique packing configuration for any given number of strands, whose effective twist radius they propose as equal to the radius of the smallest circle that circumscribes the whole packing. Gaponov \textit{et al.}\cite{Gaponov2013} take a different approach, noting that the radius of a packing could change under twist and that $r(\theta)$ be defined from Eq.~(\ref{eq:theory}) as $r(\theta)=\sqrt{(L^2-{L_0}^2)/\theta^2}$. Notably, their paper is the only one we are aware of that reports the effective twist radius of a single strand: $r_{01} \approx 0.7r_E$ where $r_E$ is the Euclidean geometrical radius.

In this paper, we measure contraction behavior for several different kinds of strings and for different numbers of strands per bundle. With this data, we highlight the extent to which Eq.~(\ref{eq:theory}) holds and we incorporate two physical effects to improve agreement with observations. The first is to invoke volume conservation, which accounts for an initially faster-than-expected contraction. The resulting increase of bundle radius with twist angle improves the range over which Eq.~(\ref{eq:theory}) holds but goes against a slower-than-expected contraction at larger twist angles. To model this, we incorporate string extension during twist, as set by the geometry and spring constant of the bundle.  This improves agreement, but does not fully account for the slower-than-expected contraction under large twist. Besides improving upon Eq.~(\ref{eq:theory}) and highlighting the need for even further modeling, our work also provides guidance on how to treat experimental data for the purpose of characterizing string and bundle behavior in terms of an effective twist radius. We finish by attempting to relate the twist radii to bundle geometry.

\section{Experimental Procedures}

\begin{figure}[h]
 \centering
 \includegraphics{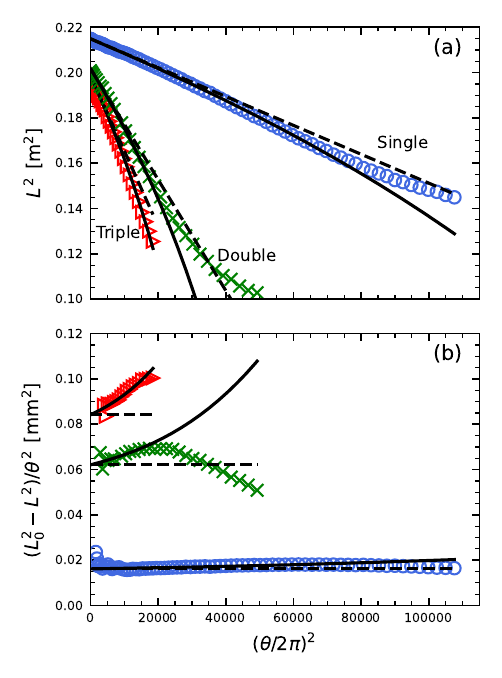}
 \caption{(a) Squared length versus squared twist angle for bundles of 0.41~mm diameter Nylon monofilament. Dashed lines represent fits of the initial contraction to Eq.~(\ref{eq:L2vQ2}), by adjusting $r_0$ and $L_0$; these correspond to the dashed curves in Fig.~\protect{\ref{fig:theorydiagram}b}. The solid curves represent Eq.~(\ref{eq:L2volexp}) with the same values of $r_0$ and $L_0$. (b) Same data and curves, replotted so that the $y$-axis represents the effective angle-dependent twist radius, squared, as defined by Eq.~(\ref{eq:theory}).}
 \label{fgr:Linearize}
\end{figure}

\begin{figure}[ht!]
\centering
\includegraphics{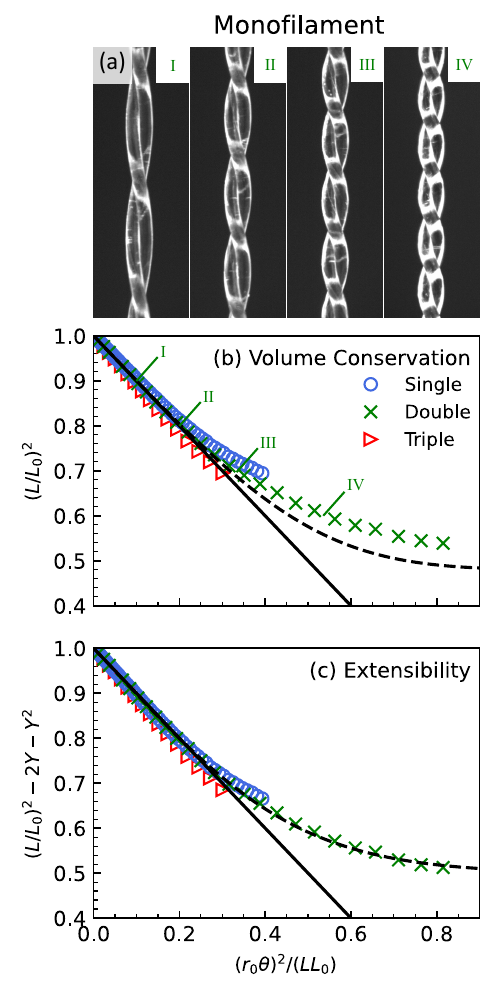}
\caption{Length vs twist angle, scaled using the initial length $L_{0}$ and fitted radius $r_{0}$, for single-, double-, and triple-strands of 0.41~mm diameter Nylon monofilament. In (a), a close up picture of the double-stranded monofilament is shown under twist at different contractions as labelled. In (b), this scaling accounts for volume conservation for a contracting cylinder, and in (c) extensibility is further considered where the correction parameter $Y$ is given by Eq.~(\ref{eq:Y}).  In both parts, the lines $y=1-x$ represents the respective expectations of Eqs.~(\ref{eq:L2vQ2VolCo},\ref{eq:L2extScaled}) while the dashed curves are the same guide to the eye, Eq.~(\ref{eq:guide}), obtained by fit to the double-stranded data in (b). }
\label{fgr:Collapsed}
\end{figure}

To critically examine the validity of Eq.~(\ref{eq:theory}) with constant twist radius, $r=r_0$, we begin with single-, double- and triple-stranded bundles of Nylon monofilament. This choice avoids complications due to the internal braid structure of woven strings. In particular, we used roughly 40~cm length bundles, with each strand having a Euclidean geometric radius of $r_E=0.205$~mm. The top of the bundle is tied around a horizontal post and constant tension is maintained by tying the bottom end to a 1~kg hanging mass. Nearly the same results were found for a range of masses, and 1~kg was safely within the tolerance of many different types of string. A horizontal wooden dowel is inserted into the bottom knot to hold the string at fixed angular displacement against two vertical posts. The string is then systematically twisted, being careful to maintain constant tension as set by the hanging mass, and its length found with a tape measure versus increasing twist until the string either approaches supercoiling or snaps.  Uncertainty in each length measurement is $0.3$~mm; all error bars are smaller than the markers on our plots, and thus not shown.  For Nylon, by contrast with the braided strings later in the paper, the material is sufficiently stiff that the cross section of each string in the bundle appears by eye to remain circular, even in the single-stranded case. Images for the double-stranded case are shown later, in Fig.~\ref{fgr:Collapsed}a.

Contraction data for all three bundles are plotted in Fig.~\ref{fig:theorydiagram}b. We see that Eq.~(\ref{eq:theory}) fits the initial contraction quite well, by adjusting $L_0$ and $r_0$, if a small-enough range of twist angles is selected; however, it noticeably deviates at larger twist angles. Better average agreement may be achieved by fitting over the whole range, as in prior works, but this gives noticeable systematic deviation across the whole range. And even more unfortunately, it gives fitting results for $r_0$ that depend on fitting range. Thus, Eq.~(\ref{eq:theory}) with constant effective twist radius $r=r_0$ can only be trusted for small twist angles, for contractions less than about 5\%.

To further highlight the deviation revealed in Fig.~\ref{fig:theorydiagram}b between Eq.~(\ref{eq:theory}) with $r=r_0$ and actual behavior, we replot the data as $L^2$ vs $(\theta/2\pi)^{2}$, since the base expectation is then a straight line of form
\begin{equation}
    L^2 = {L_0}^2 \left[ 1 - \left(\frac{r_0 \theta }{L_0}\right)^2\right].
\label{eq:L2vQ2}
\end{equation}
As seen in Fig.~\ref{fgr:Linearize}a, the data are linear only initially. At medium and large twist angles they deviate as per Fig.~\ref{fig:theorydiagram}b, but now more obviously.  This makes it easier to choose the fitting range for the initial linear regime, where the slope gives a well-defined effective radius $r_0$ for the bundle of strings.  Fit values for the effective radii are collected and discussed in a later section along with results for other string materials. For the three cases in Fig.~\ref{eq:L2vQ2}, at medium twist angles the data all fall below the prediction, \textit{i.e.} the contraction becomes faster than expected.  At even larger twist angles, for contractions greater than about 20\%, the data generally turn upwards and the contraction becomes slower than expected. This same behavior can be seen in Fig.~\ref{fgr:Linearize}b where the data and curves are plotted as $(L^2-{L_0}^2)/\theta^2$, which equals the square of the angle-dependent effective twist radius defined by solving Eq.~(\ref{eq:theory}) for $r$.  Note that experimental results for $r(\theta)$ equal $r_0$ at $\theta=0$, increase with $\theta$ at small angles, and then decrease at large angles.

\section{Models and Experimental Tests}

We hypothesize that the initial deviation from constant-$r_0$ behavior, where the contraction is faster than expected such that $r(\theta)$ initially increases with $\theta$, is due to neglect of volume conservation. Indeed, a homogeneous solid cylindrical string like Nylon monofilament must thicken if it remains cylindrical under contraction. We might expect this to hold for bundles too, as has been remarked upon previously~\cite{Gaponov2014}. To account for the conservation of string volume $V$, we allow the radius $r$ to increase with contraction according to $V = \pi r^2 L = \pi r_0^2 L_0$. Isolating $r$ and substituting into Eq.~(\ref{eq:L2vQ2}) gives
\begin{equation}
    \left(\frac{L}{L_{0}}\right)^2 = 1 - \frac{(r_{0} \theta)^{2}}{L L_{0}}.
\label{eq:L2vQ2VolCo}
\end{equation}
This cubic equation can be solved exactly for $L$; however, the result is a bit cumbersome and in principle cannot capture the slower-than-expected contraction at large twist angles.  Therefore, it suffices to consider the expansion of the exact solution:
\begin{equation}
    \left(\frac{L}{L_{0}}\right)^2 = 1 - \left(\frac{r_0\theta}{L_0}\right)^2
    -\frac{1}{2}\left(\frac{r_0\theta}{L_0}\right)^4
    -\frac{5}{8}\left(\frac{r_0\theta}{L_0}\right)^6
    -\left(\frac{r_0\theta}{L_0}\right)^8 - \ldots
\label{eq:L2volexp}
\end{equation}
This function is shown by the solid curves in Fig.~\ref{fgr:Linearize}, using the same $r_0$ and $L_0$ values as before. These curves match the initial deviation from constant-$r_0$ contraction quite well, with no additional parameters. This is also seen in Fig.~\ref{fgr:Linearize}b, where the $y$-axis represents $r^2(\theta)$, which increases from ${r_0}^2=r(0)^2$ more rapidly for the thicker bundles since all lengths are almost the same. It can also be seen in Fig.~\ref{fgr:Collapsed}b, where $(L/L_0)^2$ data are plotted versus $(r_0\theta)^2/LL_0$ and found to hug $y=1-x$ to larger twist angles, no longer falling below the line. The improved agreement up to intermediate twist angles supports our hypothesis that volume conservation is responsible for the leading correction to Eq.~(\ref{eq:theory}) with constant $r=r_0$.
We can thus recommend that Eq.~(\ref{eq:L2volexp}) be used to fit for the value of $r_0$, in order to characterize the contraction behavior of unknown bundles, because it matches data over a substantially larger range.

At larger twist angles, the contraction is generally slower than expected. We hypothesize that this deviation may be partly due to the elastic extensibility of the strings, which we model as follows. If each strand has spring constant $k$, and the bundle is held under constant tension $T$ by a hanging mass, then the untwisted length $L_0$ is greater than the relaxed length $\Tilde L_0$ according to $T=Nk(L_0 - \Tilde L_0)$. When twisted by $\theta$, the helix length $L_H$ grows even further so that the axial components of the strings' tensions sum to the bundle tension: $T=NT_s\cos\alpha$ where $T_s=k(L_H - \Tilde L_0)$ is the tension of each strand, $\alpha$ is the helix angle, $\cos\alpha=L/L_H$, and $L=\sqrt{L_H^2 - (r\theta)^2}$ is the contracted length of the bundle (all per Fig.~\ref{fig:theorydiagram}a but with hypotenuse $L_H$). In essence, the strand tensions increase as the helix angle increases in order to achieve force balance axially. Radially, the strings' individual tensions are balanced against string-string normal contact forces (as a first approximation we ignore the resulting compressibility, which must decrease the effective radius and slow the contraction rate). Further using the volume conservation condition $r_0^2L_0=r^2L$, these ingredients combine to give
\begin{equation}
    \left(\frac{L}{L_{0}}\right)^2 = \left(\frac{ 1-\frac{T}{NkL_0}}{1 - \frac{T}{NkL} }\right)^{2} - \frac{(r_{0} \theta)^{2}}{L L_{0}}
\label{eq:L2vQ2VolExtCor}
\end{equation}
Note that both sides equal one at $\theta=0$, where $L=L_0$.  Also note that this reduces to Eq.~(\ref{eq:L2vQ2VolCo}) if there is no extension ($T\to0$ or $k\to\infty)$. As before, it is straightforward to develop a series solution:
\begin{equation}
    \left(\frac{L}{L_0}\right)^2 = 1 - (1-t)\left(\frac{r_0\theta}{L_0}\right)^2
    -\frac{1}{2}(1-t)(1-5t/2)\left(\frac{r_0\theta}{L_0}\right)^4 -\ldots
\label{eq:L2vQ2VolExtCorExpansion}
\end{equation}
where $t=T/(NkL_0)$ is a dimensionless number quantifying extensibility. Note that $T=(Nk)x$ gives the fractional extension of the untwisted bundle as $x/L_0=T/(NkL_0)=t$.


In order to test this prediction, we first measure the spring constant for the nylon monofilament. By hanging a range of masses from one, two, and three stranded bundles and measuring the extension, we find $k=325\pm12$~N/m. This gives a Young's modulus of 1.1~GPa, which is within the 0.2--4~GPa range commonly quoted for Nylon materials along with a Poisson's ratio of about 0.4.  Next, we define a dimensionless parameter
\begin{equation}
    Y=\frac{\frac{T}{Nk}\left(\frac{1}{L}-\frac{1}{L_0}\right)}{1-\frac{T}{NkL}}
\label{eq:Y}
\end{equation}
for the extensibility correction to contraction under twist. This allows Eq.~(\ref{eq:L2vQ2VolExtCor}) to be rewritten as follows, such that Eq.~(\ref{eq:L2vQ2VolCo}) is recovered for $Y=0$:
\begin{equation}
    \left(\frac{L}{L_{0}}\right)^2 - 2Y - Y^{2} = 1 - \frac{(r_{0} \theta)^{2}}{L L_{0}}
\label{eq:L2extScaled}
\end{equation}
Data should thus fall on the line $y=1-x$ when experimental results for the left-hand side are plotted versus $(r_0\theta)^2/(LL_0)$. For this we use the same $L_0$ values but divide the previously-found $r_0$ values by $\sqrt{1-t}$, which according to Eq.~(\ref{eq:L2extScaled}) will ensure the same initial contraction rate. This analysis is illustrated in Fig.~\ref{fgr:Collapsed}c, where we find that data follow the expectation to slightly larger twist angle than with the volume-conservation correction alone. As an empirical guide to the eye, we also include the curve
\begin{equation}
    y=1-a\tanh\left[(x/a) + bx^2\right]
\label{eq:guide}
\end{equation}
where the parameters $a=0.5$ and $b=0.6$ were obtained by fitting to the double-strand case. This matches the data better than using $+bx^3$ inside the $\tanh()$ function, though it might be preferable because $+bx^2$ gives a negative leading-order correction to $y=1-x$.

\begin{figure}[ht!]
    \includegraphics[width=8.5cm,height=8.5cm,keepaspectratio]{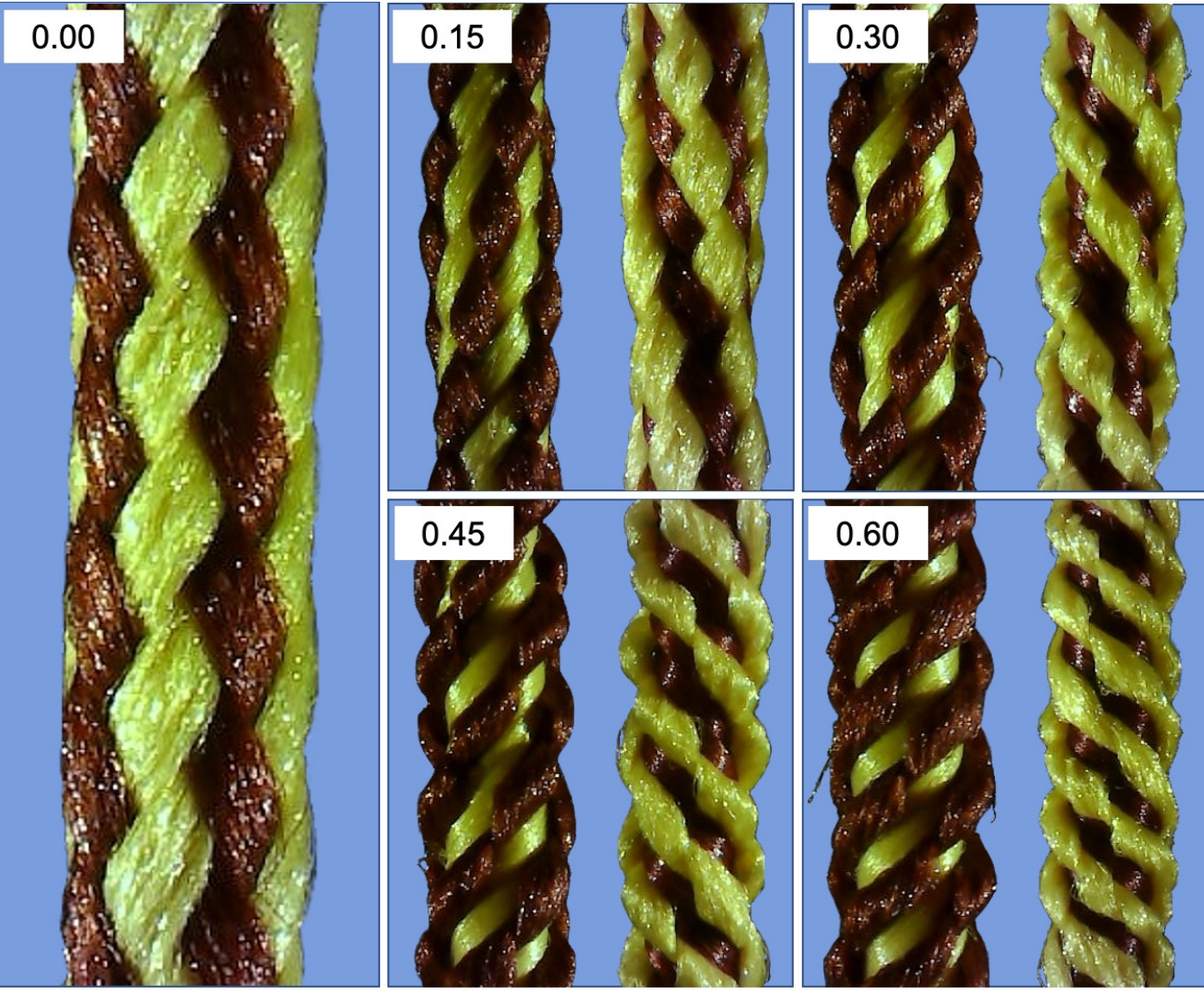}
    \caption{A single strand of nylon boot lace with 1.9~mm geometrical radius and with brown and yellow components woven together in opposite chirality. Photographs are for the untwisted case (left), as well as four different amounts of clockwise and counterclockwise twist; inset numbers refer to the scaled twist angle squared, $(r_0\theta /LL_{0})^{2}$. In a given direction of twist, one color component tenses while the other relaxes and bows outward. When twisted in the opposite direction, the color components swap behavior. }
    \label{fig:TwistedStrings}
\end{figure}

While our modeling attempts appear to capture physical effects, there remains considerable deviation at high twist angles where the contraction is still slower than expected. We can only speculate on potential reasons, which could be addressed by future studies. As a first possibility, the force balance conditions assumed for modeling extensibility did not consider static strand-strand friction, which could be mobilized as the bundle is twisted. However, this effect would be absent for the single-stranded monofilament (though not for the single-stranded woven strings in the next section). A second explanation could be the deformation of individual strands away from circular cross-section due to compressive normal contact forces. For multiple strands this is induced by compressive strand-strand contact forces, mentioned parenthetically above, as well as by the spatial distribution of stresses throughout the volume of each strand.  For single strands, the latter is illustrated in Fig.~\ref{fig:TwistedStrings} by photographs of a woven nylon boot lace, where two color components are braided together with opposite chirality such that the lace itself remains achiral. Evidently, under twist, this lace does not uniformly expand. Rather, woven components of the same chirality as the twist direction contract while the opposite components bow outward, forming a pronounced helical structure. The inner one is clearly taut, while the other is relaxed. A similar helical structure actually becomes noticeable for a Nylon monofilament under sufficient twist, but is less pronounced and difficult to photograph. Such structures belie a non-circular cross section and a nontrivial pattern of elastic stresses within the material.

\section{Various String Materials and Braids}

Using the same data collection and analysis procedures as above, we now characterize the contraction behavior under twist for woven strings of six different braid structures and materials.  This includes two kinds of nylon rattail cord, cording nylon, and Kevlar (all of which are single-component and achiral), as well as parachute cord (which has an inner core and outer sheath) and yarn (which has a chiral structure). Due to the loose / fuzzy / floofy braid structure, the Euclidean geometrical radii $r_E$ are ill-defined to varying degrees. Also by contrast with Nylon, the cross section of individual strands deforms under twist - but is beyond our scope to measure. Results for our best efforts to measure $r_E$, as well as less uncertain results for the linear mass densities $\lambda$ and spring constants $k$, are collected in Table~\ref{config_info}.

\begingroup
\setlength{\tabcolsep}{.18cm} 
\renewcommand{\arraystretch}{2} 
\begin{table*}[ht] 
\begin{center}
\begin{tabular*}{\textwidth}{@{\extracolsep{\fill}}lllllll}
\hline
\hline 
\textbf{Material} & ${r_E}$ (mm) & $\lambda$ (g/m)  &  $k$ (N/m) & $r_{01}$ (mm) & $r_{02}$ (mm) & $r_{03}$ (mm) \\

\hline
Monofilament  & 0.205 & 0.058 $\pm$ 0.001 & 325 $\pm$ 10 & 0.1321 $\pm$ 0.0002 & 0.2488 $\pm$ 0.0008 & 0.285 $\pm$ 0.001 \\

Silver Rattail & 0.89 & 1.463 $\pm$ 0.002 & 760 $\pm$ 40 & 0.522 $\pm$ 0.002 & 0.855 $\pm$ 0.002 & 1.081 $\pm$ 0.002 \\

Rainbow Rattail & 0.93 & 1.482 $\pm$ 0.002 & 600 $\pm$ 30 & 0.561 $\pm$ 0.001 & 0.855 $\pm$ 0.002 & 1.190 $\pm$ 0.002\\
Cording Nylon & 0.55 & 0.470 $\pm$ 0.004 & 380 $\pm$ 10 & 0.264 $\pm$ 0.001 & 0.383 $\pm$ 0.001 & 0.528 $\pm$ 0.001\\
Kevlar & 0.44 &0.464 $\pm$ 0.004  & 2300 $\pm$ 100& 0.502 $\pm$ 0.002 & 0.702 $\pm$ 0.002 & 0.814 $\pm$ 0.002 \\ 
Parachute & 1.07 &4.200 $\pm$ 0.004 & 2350 $\pm$ 30& 1.36 $\pm$ 0.01 & 1.67 $\pm$ 0.01 & 1.93 $\pm$ 0.01 \\
Yarn & 1.7 $\pm$ 0.1 & 0.666 $\pm$ 0.003 & 255 $\pm$ 10 & 0.29 $\pm$ 0.03 & 0.53 $\pm$ 0.03 & 0.67 $\pm$ 0.03 \\
\hline
\hline
\end{tabular*}
\end{center}
\caption{Physical parameters for the various string materials and bundles.  The Euclidean geometric radius $r_E$, linear mass density $\lambda$, and spring constant $k$ for single strands were all measured directly. The twist radii $r_{0N}$ for $N$-stranded bundles were obtained by fits to initial behavior for contraction versus twist angle. Additionally, the maximum twist angles prior to supercoiling were in the range $(0.8-0.9)L_o/r_{0N}$, and the corresponding maximum contractions were in the range $(20-30)\%$, in all cases.}
\label{config_info}
\end{table*}
\endgroup

\begin{figure}[ht!]
\centering
\includegraphics{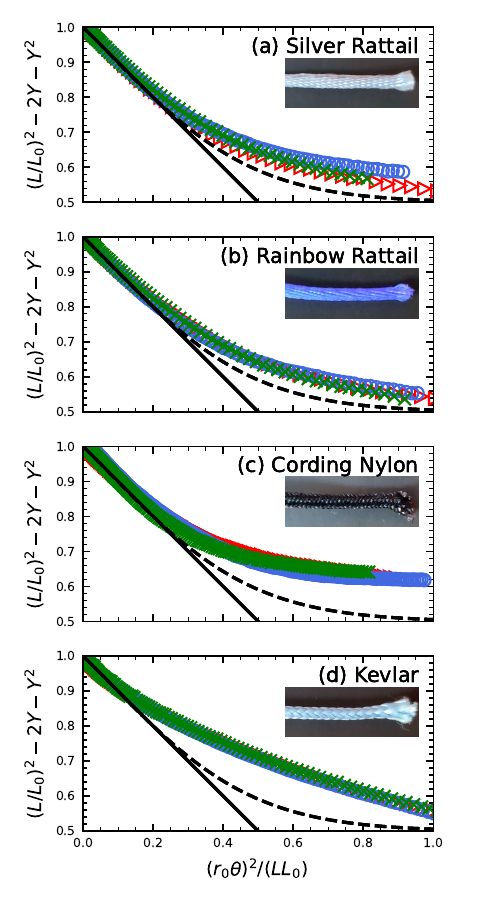}
\caption{Squared length vs squared twist angle, scaled by initial length $L_{0}$ and fitted radius $r_{0}$, for (a) silver colored rattail, (b) rainbow colored rattail, (c) cording Nylon, and (d) Kevlar strings. Each string is made of a different material, and all but the two rattail strings have different braid structures. For each string, single-, double-, and triple-stranded strings are plotted as circles, crosses, and triangles, respectively. The solid black line shows the leading order linear behavior, while the dashed curves are the same guide to the eye used for the monofilament, Eq.~(\ref{eq:guide})  }
\label{fgr:AllBraid}
\end{figure}

In Fig.~\ref{fgr:AllBraid} we display contraction data for the four strings that are achiral and relatively uniform, and thus expected to be more ideal. As before, data are collected up to twist angles where the bundle either snaps or is about to supercoil. The volume and extensibility corrections are included, so that the data hug the line $y=1-x$ for greater amounts of twist and enable better measurement of the effective twist radii. The resulting values, collected in Table~\ref{config_info}, are consistent with fits to Eq.~(\ref{eq:theory}) at smaller angles. But in all cases, the deviation from linearity in Fig.~\ref{fgr:AllBraid} is a slower-than-expected contraction that is more pronounced in comparison with the monofilament case as judged against same guide-to-the-eye curve shown in Fig.~\ref{fgr:Collapsed}. Note that this nonideality is greatest for Kevlar and smallest for the rattails, and hence may correspond with internal friction. Also note that in all four plots the data for different numbers of strands collapse fairly well onto a materials-dependent curve, perhaps even better than for the monofilament. This collapse lends further support to our analysis procedure. Interestingly, for the cording Nylon, the collapsed data asymptotes toward a constant corresponding to a halt of contraction at large twist angles. This motivated our earlier choice of a hyperbolic tanh function as a guide-to-the eye for the deviation from linearity.

\begin{figure}[ht]
\includegraphics{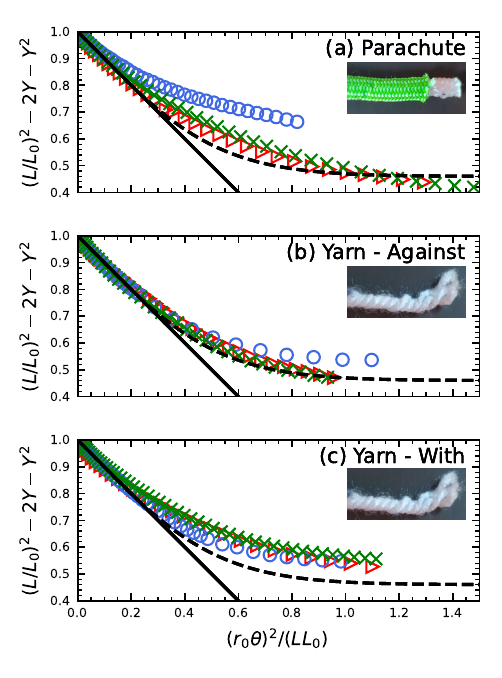}
\caption{Squared length vs squared twist angle, scaled by initial length $L_{0}$ and fitted radius $r_{0}$, for (a) parachute cord with a central core, (b) twisting yarn against its chirality, and (c) twisting yarn with its chirality. For each string, single-, double-, and triple-stranded strings are plotted as circles, crosses, and triangles, respectively. The solid black line shows the leading order linear behavior, while the dashed curves are the same guide to the eye used for the monofilament, Eq.~(\ref{eq:guide})  }
\label{fgr:Oddities}
\end{figure}

Out of curiosity we also investigated three cases where the string structure is more intricate, shown in Fig.~\ref{fgr:Oddities}. The parachute cord (Fig.~\ref{fgr:Oddities}a) is composed of two entirely separate concentric structures, a central core that is freely surrounded by an outer sheath, each with a woven braid. The yarn (Fig.~\ref{fgr:Oddities}b and Fig.~\ref{fgr:Oddities}c) has a chiral braid, and so its behavior is dependent on the direction of twist. For the parachute and unwinding yarn cases, the same high degree of collapse is not found. The internal structure of the strings alters their behavior in the single-stranded case compared to the double- and triple-stranded cases. As the number of strands increases, we see the data begins to collapse, suggesting the behavior is dominated by the inter-strand interaction rather than the individual strand behavior. We also see this difference disappears in the yarn if we twist in the same direction as its chirality (Fig.~\ref{fgr:Oddities}c).

\begin{figure}[h]
    \centering
    \includegraphics[width=8.5cm]{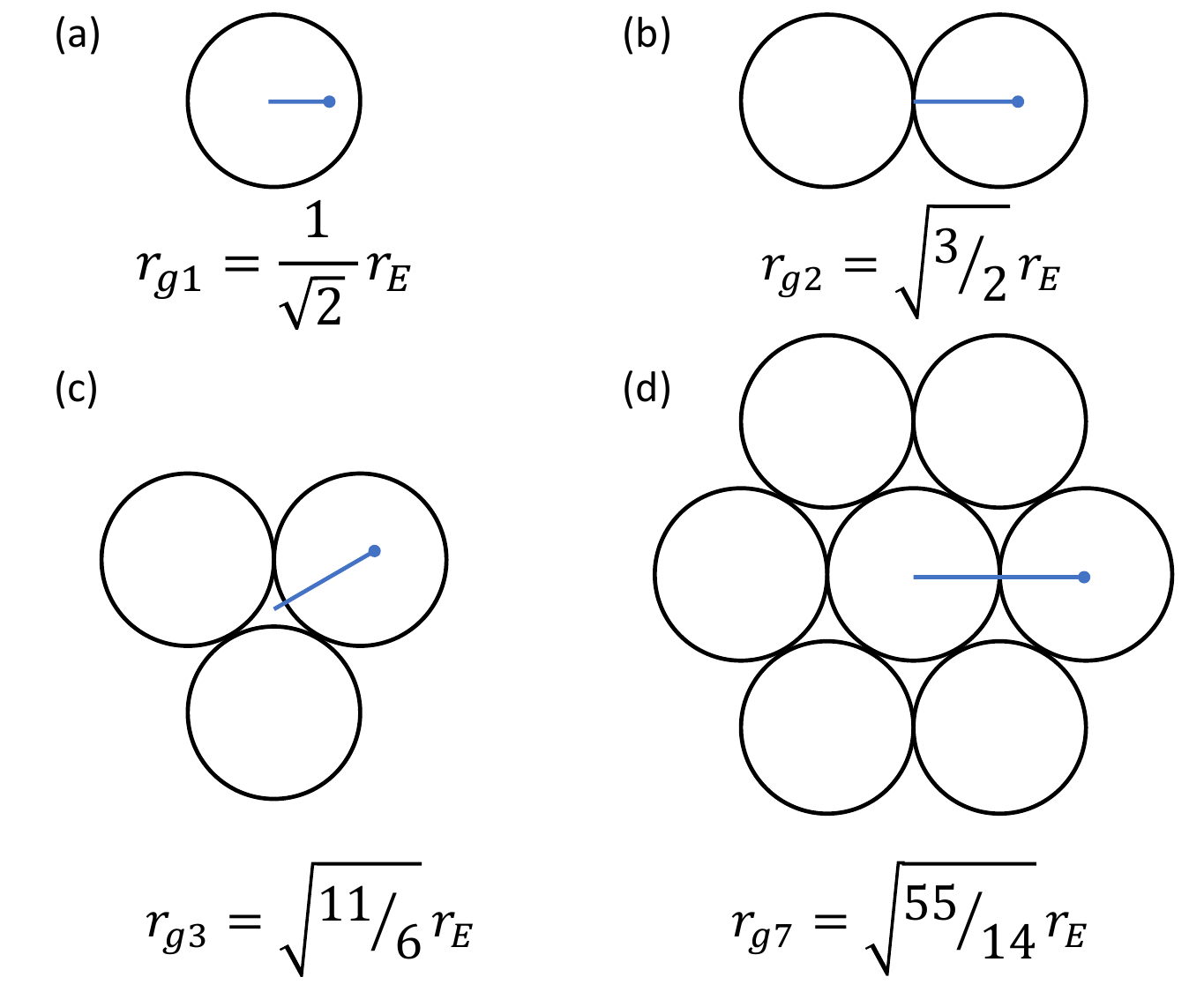}
    \captionof{figure}{Radius of gyration $r_{gN}$ for different number $N$ of strands within a bundle in terms of the Euclidean geometric radius $r_{E}$ of a single strand. The expectation for the effective radius of a bundle can be described by assuming there is an internal core that does not contribute significantly to the twisting\cite{Wurtz2010}. We calculate this value instead using the radius of gyration, and find good agreement in numerical values for the monofilament case. }
    \label{fgr:GeomExp}
\end{figure}

\section{Effective Twist Radii}

The effective twist radii collected in Table~\ref{config_info} serve to describe the contraction behavior for small twist angles via Eq.~(\ref{eq:theory}), and for somewhat larger twist angles via Eq.~(\ref{eq:L2vQ2VolExtCor}). But what sets their values? We again focus on the monofilament results to build intuition. Whereas the other strings in this paper present idiosyncratic behavior based on braid and material properties, the monofilament retains a well-defined geometrical structure and has the strongest adherence to ideality. As outlined in the introduction, much of the literature suggests that radius of contraction in the double-stranded case, $r_{2}$, should equal the Euclidean radius, $r_{E}$. For the monofilament we used, $r_{E}=0.205$~mm, however, our fit value $r_{02}=0.249$~mm is larger by more than $20\%$. To summarize our fitting results for twist radii, we find empirically that the radius of gyration is a useful length scale. The radius of gyration defines the radius of a ring with the same mass and moment of inertia as an object of arbitrary dimension, and is illustrated for single-, double-, triple-, and seven-stranded bundles in Fig.~\ref{fgr:GeomExp}. This length scale could extend the notion in literature of a radius below which strands do not contribute to twisting by continuously defining that threshold. We also find excellent agreement in experiment; for the double-stranded monofilament $r_{g2} = 1.22r_{E} = 0.251$~mm a close match to our fit value of $r_{02}=0.248$~mm. This formulation also allows us to extend a prediction to the single-stranded case, where $r_{g1} = 0.707r_{E} = 0.145$~mm, within $.01$~mm of the value we found. Interestingly, Gaponov \textit{et al.}\cite{Gaponov2013} noted this same phenomenon, finding the single stranded radius is $\approx 0.7$ times the geometric radius, matching the prediction from the radius of gyration.

\begin{figure}[h]
\includegraphics{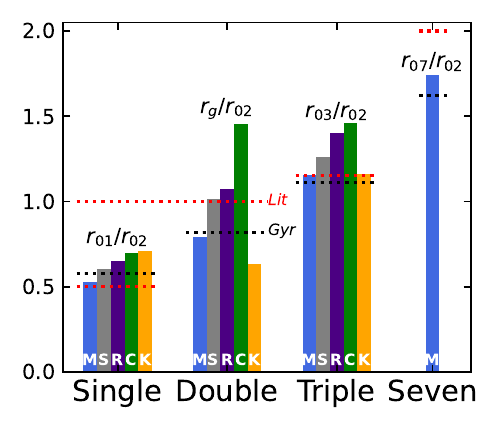}
\caption{Fit and putative radii scaled by double-stranded fit radius, $r_{02}$. White label indicates the braid structures (monofilament (M), rattail (S and R), cording Nylon (C), Kevlar (K), and parachute cord (P)) and material (silver rattail (S) vs rainbow rattail (R)) shown. Strings are grouped by number of strands used in the bundle, with the double-stranded case instead comparing against the putative radius, $r_{E}$, measured optically in the taut but untwisted case. Dashed black lines indicate expectation from the radius of gyration, while dashed red lines indicate expectation from literature. Strings are ordered left to right in descending order of maximum scaled arc length through which the string contraction remains linear. We observe in the single stranded case, the deviation from expectation of the scaled radius, $r_{01}/r_{02}$, also corresponds to the deviation from linearity.}
\label{fgr:Radii}
\end{figure}

The monofilament case, having no internal braid structure, allows us to confidently define the geometrical radius. The rest of the strings we examined have complicated braid structures; this leads to an ambiguous definition of a strand radius based on string packing, material parameters and as yet unidentified variables. However, we can still make predictions about the ratios of radii between bundles. In Fig.~\ref{fgr:Radii} we plot these ratios, normalizing all values by the double-stranded case as this is the most common case described in literature. From left to right in this figure, we show the single-stranded radius, the geometrical radius, and the triple-stranded radius respectively. Looking first at the geometrical radius (center plot), we observe what was described above. The dashed red line indicates the expectation found in literature, $r_{2} = r_{E}$ and the dashed black line for the expectation from the radius of gyration, $r_{2} = 1.22 r_{E}$. The monofilament agrees strongly, with variance between the other strands. We see an interesting disagreement between the silver and rainbow rattails, two strings with identical braid structure but slightly different materials. We also see that braid structure does not only cause a decrease in the double-stranded radius, as the Kevlar has an even larger effective radius than expected from the radius of gyration. Under twist we may expect a string to compress to fill space between strands or expand to conserve volume. As highlighted in Fig.~\ref{fig:TwistedStrings}, it may do both as chiral strands in the opposing direction of twist flare out to compensate the strands under tension. The net effect in the silver, rainbow and cording nylon cases seems to be contraction to a smaller than expected radius, while the Kevlar expands.

Using the single- and triple-stranded cases, we highlight two features: First, looking at the monofilaments, the geometrical expectation from literature seems a better match for ratio between strands than the radius of gyration. This would suggest, once a proper effective radius has been defined, perhaps the unbraided structure does behave closer to ideality in a geometric packing. Second, across the braid structures examined in this paper, we see a hierarchy of deviation from that ideal single-to-double-stranded ratio. This deviation from the ideal ratio coincides with the deviation from the linear regime in Fig. \ref{fgr:AllBraid}, with the silver rattail remaining linear over the largest twist range and Kevlar deviating at very low twist angle. The same is not seen in the triple-stranded case, where the Kevlar instead behaves almost ideal to literature expectation. Perhaps the deviation of single-strand behavior has strong bearing on the onset of non-linear behavior, even if those strands may pack geometrically at larger bundle numbers.

\section{Conclusions}

In this paper we presented contraction versus twist angle data for a wide range of string materials and for different numbers of strands in a bundle, and we presented ways to critically compare observations against the standard model, Eq.~(\ref{eq:theory}). We find two distinct types of deviation common to all cases.  At medium twist angles, for contractions in the range of about 5--20\% varying by string, the contraction is faster than expected.  This effect is successfully modeled using volume conservation, where the bundle thickens as it contracts, and is potentially important for reducing fitting range-dependent systematic errors when extracting an effective twist radius from experimental data. Following on from this, we tabulated the effective twist radii for the range of systems and found that they do not compare as well with geometrical radii as suggested by previous investigations.  Rather, they compare better with the radius of gyration. Second, at large twist angles, for contractions beyond about 15\%, the contraction is slower than expected. We uncovered some hints that this is due to elastic effects, both in the extension of the individual strings in a bundle but also in compressional deformation due to string-string normal contact forces.  We modeled the former, but not the latter. This lack of full understanding, and the empirical but not yet theoretically-justified findings about the values of the effective twist radii, may hopefully inspire further theoretical and experimental work.

\section*{Conflicts of interest}
There are no conflicts to declare.

\section*{Acknowledgements}
This work was supported by NSF grants REU-Site/DMR-1659515 and MRSEC/DMR-1720530.  GED also thanks the Meyerhoff Scholars Program and the NOAA Educational Partnership Program (contract numbers NA11SEC481004.3 and NA16SEC4810008) for their support.



\bibliography{BuzzerRefs.bib} 
\bibliographystyle{rsc.bst.txt} 

\end{document}